%% file: 0masktof_main.tex
\documentclass[10pt,twocolumn,letterpaper]{article}

\usepackage{cvpr}
\usepackage{times}
\usepackage{epsfig}
\usepackage{graphicx}
\usepackage{amsmath}
\usepackage{amssymb}

\usepackage{xfrac}
\usepackage{subcaption}
\usepackage[export]{adjustbox}
\usepackage[table]{xcolor}
\usepackage{color}
\usepackage{rotating}
\usepackage{array}
\usepackage{fixltx2e}

\definecolor{myblue}{rgb}{.1,0.3,0.9}
\definecolor{rowblue}{RGB}{220,230,240}%{230,240,250}
\usepackage{booktabs}
\usepackage{tablefootnote}
\usepackage{colortbl}
\usepackage{tabularx}
\usepackage{multirow}

\usepackage[pagebackref=true,breaklinks=true,letterpaper=true,colorlinks,bookmarks=false]{hyperref}

\cvprfinalcopy % *** Uncomment this line for the final submission

\def\cvprPaperID{5463} % *** Enter the CVPR Paper ID here

\ifcvprfinal\fi
\begin{document}
% \setlength\abovedisplayskip{0.9em}
% \setlength\belowdisplayskip{0.9em}
% \setlength\baselineskip{0.7em}

%%%%%%%%% TITLE
\title{Mask-ToF:~ Learning Microlens Masks for \\ Flying Pixel Correction in Time-of-Flight Imaging}
\vspace{-15pt}
\author{Ilya Chugunov$^1$\qquad  Seung-Hwan Baek$^1$\qquad  Qiang Fu$^2$ \qquad Wolfgang Heidrich$^2$  \qquad Felix Heide$^1$ \vspace{5pt}\\
$^1$Princeton University \quad $^2$King Abdullah University of Science and Technology}

\maketitle

%\footnotetext[1]{Shared contribution. See Author Contributions.}

% Remove page # from the first page of camera-ready.
\ifcvprfinal\thispagestyle{empty}\fi

\input{definition}

%%%%%%%%% ABSTRACT
\vspace{-15pt}
\begin{abstract}
We introduce Mask-ToF, a method to reduce flying pixels (FP) in time-of-flight (ToF) depth captures. FPs are pervasive artifacts which occur around depth edges, where light paths from both an object and its background are integrated over the aperture. This light mixes at a sensor pixel to produce erroneous depth estimates, which can adversely affect downstream 3D vision tasks. Mask-ToF starts at the source of these FPs, learning a microlens-level occlusion mask which effectively creates a custom-shaped sub-aperture for each sensor pixel. This modulates the selection of foreground and background light mixtures on a per-pixel basis and thereby encodes scene geometric information directly into the ToF measurements. We develop a differentiable ToF simulator to jointly train a convolutional neural network to decode this information and produce high-fidelity, low-FP depth reconstructions. We test the effectiveness of Mask-ToF on a simulated light field dataset and validate the method with an experimental prototype. To this end, we manufacture the learned amplitude mask and design an optical relay system to virtually place it on a high-resolution ToF sensor. We find that Mask-ToF generalizes well to real data without retraining, cutting FP counts in half.
\end{abstract}

%%%%%%%%%% BODY TEXT

\vspace{-14pt}
\input{1introduction}

%------------------------------------------------------------------------
\input{2related_work}
%------------------------------------------------------------------------
\input{3image_formation}
%------------------------------------------------------------------------
\input{4method}

%------------------------------------------------------------------------
\input{5synthetic_assessment}

%------------------------------------------------------------------------
\input{6experimental_assessment}
%------------------------------------------------------------------------
\input{7conclusion}

%------------------------------------------------------------------------
\vspace{2pt}

{\small
\bibliographystyle{ieee_fullname}
\bibliography{bib}
}

\end{document}

%% file: definition.tex
% formatting stuff

\definecolor{Gray}{rgb}{0.5,0.5,0.5}
\definecolor{darkblue}{rgb}{0,0,0.7}
\definecolor{orange}{rgb}{1,.5,0} % something readable but different from todo
\definecolor{red}{rgb}{1,0,0} % something readable but different from todo

% taken from https://designnavigator.daimler.com/Daimler_Color_System
\definecolor{dai_ligth_grey}{RGB}{230,230,230}
\definecolor{dai_ligth_grey20K}{RGB}{200,200,200}
\definecolor{dai_ligth_grey40K}{RGB}{158,158,158}
\definecolor{dai_ligth_grey60K}{RGB}{112,112,112}
\definecolor{dai_ligth_grey80K}{RGB}{68,68,68}
\definecolor{dai_petrol}{RGB}{0,103,127}
\definecolor{dai_petrol20K}{RGB}{0,86,106}
\definecolor{dai_petrol40K}{RGB}{0,67,85}
\definecolor{dai_petrol80}{RGB}{0,122,147}
\definecolor{dai_petrol60}{RGB}{80,151,171}
\definecolor{dai_petrol40}{RGB}{121,174,191}
\definecolor{dai_petrol20}{RGB}{166,202,216}
\definecolor{dai_deepred}{RGB}{113,24,12}
\definecolor{dai_deepred20K}{RGB}{90,19,10}
\definecolor{dai_deepred40K}{RGB}{68,14,7}
%\definecolor{violettblau}{cmyk}{0.9, 0.6, 0, 0}
\definecolor{rot}{RGB}{238, 28 35}
\definecolor{apfelgruen}{RGB}{140, 198, 62}
%\definecolor{gelb}{RGB}{255, 229, 0}
\definecolor{orange}{RGB}{244, 111, 33}
\definecolor{pink}{RGB}{237, 0, 140}
\definecolor{lila}{RGB}{128, 10, 145}
%\definecolor{hellgrau}{RGB}{224, 224, 224}
%\definecolor{mittelgrau}{RGB}{128, 128, 128}
%\definecolor{dunkelgrau}{RGB}{80,80,80}
\definecolor{anthrazit}{RGB}{19, 31, 31}

\newcommand{\heading}[1]{\noindent\textbf{#1}}
\newcommand{\note}[1]{{\em{\textcolor{orange}{#1}}}}
\newcommand{\todo}[1]{{\textcolor{red}{\bf{TODO: #1}}}}
\newcommand{\comments}[1]{{\em{\textcolor{orange}{#1}}}}
\newcommand{\changed}[1]{#1}
\newcommand{\place}[1]{ \begin{itemize}\item\textcolor{darkblue}{#1}\end{itemize}}
\newcommand{\de}{\mathrm{d}}

\newcommand{\normlzd}[1]{{#1}^{\textrm{aligned}}}

% dimensions
\newcommand{\ttime}{\tau}               % time coordinate
\newcommand{\x}{\Vect{x}}               % spatial coordinates in vectorized form
\newcommand{\z}{z}               % depth coordinate of volume

\newcommand{\NEW}[1]{{\color{blue}{#1}}} % for writing
\newcommand{\OLD}[1]{{\color{gray}{#1}}}% for revision

\newcommand{\BAEK}[1]{{\color{orange}{\{BAEK:#1\}}}}% for comments
\newcommand{\ILYA}[1]{{\color{darkblue}{\{ILYA:#1\}}}}% for comments
\newcommand{\FELIX}[1]{{\color{purple}{\{FELIX:#1\}}}}% for comments

\newcommand{\npixels}{n}               % num pixels
\newcommand{\ntime}{t}               % num timesteps

%image formation
\newcommand{\illfunc}     {g}
\newcommand{\pathfunc}     {s}
\newcommand{\camfunc}     {f}

% notation for image formation
\newcommand{\irradiance}{E}
\newcommand{\exposure}{b}
\newcommand{\pmdfunc}{f}                % modulation on the PMD side
\newcommand{\lightfunc}{g}              % modulation of the light
\newcommand{\period}{T}                 % temporal period of the modulation
\newcommand{\freqm}{\omega}                % frequency of modulation
\newcommand{\illphase}{\rho}             % frequency of modulation harmonics
\newcommand{\sensphase}{\psi}             % frequency of modulation harmonics of pmd camera
\newcommand{\pmdphase}{\phi}            % phase of PMD modulation
\newcommand{\omphi}{{\omega,\phi}}      % shorthand for freq/phase pair
\newcommand{\numperiod}{N}              % #periods integrated over
\newcommand{\att}{\alpha}               % geometric & photometric attenuation
\newcommand{\pathspace}{{\mathcal{P}}}  % space of all light paths

% regular modulation based cameras
\newcommand{\atan}{\operatorname{atan}}

% math stuff
\newcommand{\Fourier}{\mathfrak{{F}}}         % fourier transform
\newcommand{\conv}     {\otimes}
\newcommand{\corr}     {\star}
\newcommand{\Mat}[1]    {{\ensuremath{\mathbf{\uppercase{#1}}}}} %Matrix
\newcommand{\Vect}[1]   {{\ensuremath{\mathbf{\lowercase{#1}}}}} %Vector
\newcommand{\Id}				{\mathbb{I}} %Identity matrix
\newcommand{\Diag}[1] 	{\operatorname{diag}\left({ #1 }\right)} %Diagonalized matrix
\newcommand{\Opt}[1] 	  {{#1}_{\text{opt}}} %Optimal point of an optimization
\newcommand{\CC}[1]			{{#1}^{*}} %Convex conjugate
\newcommand{\Op}[1]     {\Mat{#1}} %Operator
\newcommand{\minimize}[1] {\underset{{#1}}{\operatorname{argmin}} \: \: } %Minimize w.r.t.
\newcommand{\maximize}[1] {\underset{{#1}}{\operatorname{argmax}} \: \: } %Maximize w.r.t.
\newcommand{\grad}      {\nabla}

% notation for method
\newcommand{\Basis}{\Mat{H}}         		% Matrix basis
\newcommand{\Corr}{\Mat{C}}             % measurement matrix
\newcommand{\correlem}{\bold{c}}             % measurement matrix element
\newcommand{\meas}{\Vect{b}}            % measurement vector
\newcommand{\Meas}{\Mat{B}}            % measurement matrix
\newcommand{\MeasNormalized}{\Mat{B}^{\textrm{new}}}            % measurement matrix
\newcommand{\Img}{H}                    % transient image
\newcommand{\img}{\Vect{h}}             % vectorized image
\newcommand{\latentresponse}{\alpha}

\newenvironment{customlegend}[1][]{%
        \begingroup
        % inits/clears the lists (which might be populated from previous
        % axes):
        \csname pgfplots@init@cleared@structures\endcsname
        \pgfplotsset{#1}%
    }{%
        % draws the legend:
        \csname pgfplots@createlegend\endcsname
        \endgroup
    }%

    % makes \addlegendimage available (typically only available within an
    % axis environment):
    \def\addlegendimage{\csname pgfplots@addlegendimage\endcsname} 

%% file: 1introduction.tex
\section{Introduction}
Large-scale image datasets such as ImageNet~\cite{deng2009imagenet} and CIFAR~\cite{cifar-10, cifar-100}, in tandem with a boom in computational resources, drastically reshaped the field of image processing. In the depth domain, a similar trend \cite{silberman2012indoor,     chang2015shapenet, dai2017scannet} has recently made the mass-acquisition of high-quality depth maps a vital prerequisite for a range of 3D graphics and vision applications. These include human-centered tasks such as pose tracking~\cite{shotton2011real, kolb2010time}, action recognition~\cite{jhuang2013towards, presti20163d}, and facial analysis~\cite{rossler2019faceforensics++}, as well as scene-understanding problems including mapping~\cite{gupta2017cognitive}, segmentation~\cite{dai2018scancomplete}, and object reconstruction~\cite{tulsiani2018multi, choy20163d, yan2016perspective}. While methods look to captured depth datasets for ground truth, the devices used to capture them are subject to a slew of error sources which, if not addressed, can hurt task performance and generalizability.

\begin{figure}[t!]
    \centering
    \includegraphics[width=1\columnwidth]{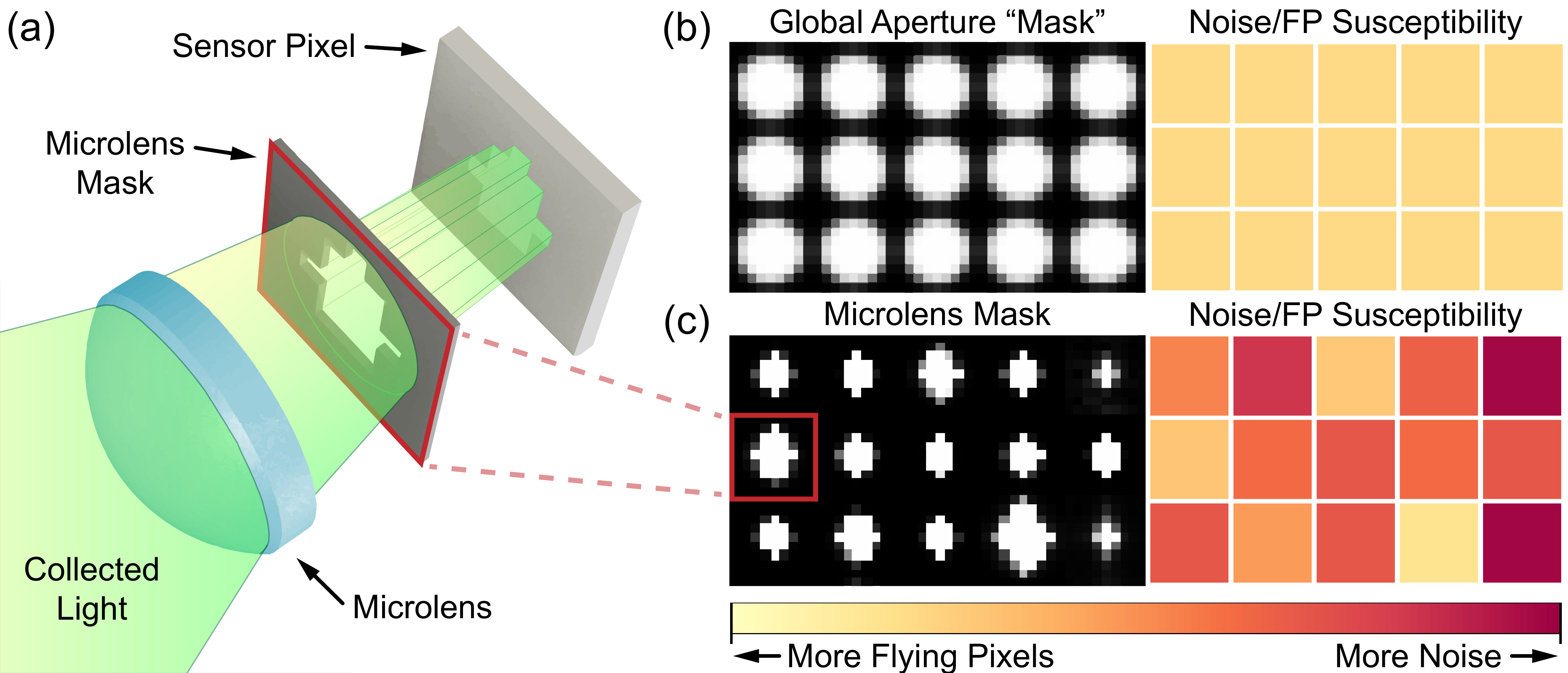}
    \vspace*{-1em}
    \caption{\label{fig:snr_fp_mini}(a) 3D visualization of a microlens mask selectively blocking light entering a sensor pixel. (b) The equivalent mask pattern for a global aperture setup, all sensor pixels equally susceptible to FPs. (c) A learned mask pattern with spatially multiplexed noise/FP susceptibility.}
    \vspace*{-1em}
\end{figure}
One of many approaches to depth acquisition is passive sensing: exploiting parallax cues to infer distances solely from input monocular~\cite{subbarao1994depth, mahjourian2018unsupervised, gordon2019depth} or multiview~\cite{hirschmuller2005accurate, kendall2017end, xu2020aanet} images. These methods can use standard RGB cameras for data acquisition, but struggle with textureless regions and complex geometries~\cite{smolyanskiy2018importance,lazaros2008review}.
Active sensing approaches tackle this challenge by first sending out a known illumination into the scene and reconstructing depth with the help of the returned light. These include structured light methods such as active stereo, where spatially patterned light is projected into the scene to aid in the stereo feature matching process~\cite{ahuja1993active}. While being robust to textureless scenes, their accuracy is fundamentally limited by illumination pattern density and sensor baseline, resulting in a bulky camera form-factor. Some of the most successful active depth sensing methods are time-of-flight (ToF) approaches, where depth is estimated from the travel time of light leaving and returning to the device. Direct ToF systems such as LiDAR send out individual laser pulses and measure their time to return using time-resolved sensors such as avalanche photo diodes~\cite{cova1996avalanche}. These can provide high-accuracy and long-range depth estimates, but use a scanning approach to collect data, leading to poor depth completeness and/or expensive sensor array systems. In contrast, amplitude-modulated continuous wave (AMCW) ToF cameras, the focus of this paper, \emph{flood-illuminate} a scene with periodic amplitude-modulated light and estimate the phase shift of returned light to infer depth. These devices do not need to time-resolve captured light like their direct ToF counterparts, and so can rely on an easy-to-manufacture CMOS sensor array to produce complete depth maps at a high framerate. This, when combined with their small sensor-illumination baseline, makes AMCW ToF cameras compact and affordable, and has led to their widespread adoption in the vision community. Devices such as those in the Microsoft Kinect series have subsequently helped create community-made freely-available scene understanding benchmarks that lower the barrier of entry for 3D vision research~\cite{song2015sun, ammirato2017dataset}.

Although they promise to democratize low-cost dense depth imaging, AMCW ToF methods are still subject to fundamental limitations of the sensing process: noise from ambient light, photon shot, phase wrapping, multipath interference (MPI), and flying pixels (FPs)~\cite{falie2007noise}. There has accordingly arisen a large body of work in computational post-processing methods to address these issues; methods concerning depth denoising~\cite{frank2009denoising, yan2018ddrnet}, phase unwrapping~\cite{lawin2016efficient}, and MPI correction~\cite{marco2017deeptof}. Contrastingly, while confidence-based methods~\cite{reynolds2011capturing} are able to identify flying-pixels, rectifying them --- recovering the depth of their corresponding chief ray --- has remained a great challenge.

FPs are formed when light from both an object and its background reaches the same sensor pixel, generating a mixed depth measurement. These often appear to be floating in empty space in the resultant point cloud, hence \textit{flying} pixels. Computationally unmixing these FPs often leads to edge blur or severe artifacts~\cite{xiao2015defocus}. As they originate in the optical pipeline, artifacts of the light collection process by the main lens, we argue that an effective strategy to mitigate them should also start in the optical pipeline. Unfortunately, a direct masking approach, such as simply reducing aperture size to block stray light paths, is not efficient for overall light throughput, and so significantly lowers SNR.

With Mask-ToF we learn a microlens amplitude mask, allowing us to generate per-pixel aperture configurations with spatially-varying susceptibility to noise and FPs, as shown in Figure \ref{fig:snr_fp_mini}. We train an encoder-decoder network which learns to aggregate this spatial information and leverages mask structural cues to produce refined depth estimates. We then backpropagate this net's loss to jointly learn high-level mask patterns. We photolithographically manufacture the learned mask, and virtually place it on the sensor with a custom optical relay system to validate Mask-ToF on real-world data. In the future, we expect this mask can be fabricated directly on the camera sensor in a similar manner to a polarization sensor~\cite{mihoubi2018survey}, preserving its form factor.

 In summary, we make the following contributions:
 
\begin{itemize}
    \item We develop a differentiable AMCW ToF image formation model, including sub-aperture light transport. \pagebreak
    \item We incorporate sub-aperture masking and a refinement network into this framework and learn an optimal mask structure through a patch-based gradient descent approach from synthetic data. \vspace{-0.5em}
    \item We test the masks in simulation, evaluating on overall error and FP reduction, then manufacture them and construct an experimental setup to validate the proposed method on real data. 
\end{itemize}

%% file: 2related_work.tex
\section{Related Work}
%-------------------------------------------------------------------------
\vspace{0.5em}\noindent\textbf{Depth Imaging.}\hspace{0.1em}
There exists a wide body of work in both passive and active methods for depth imaging. The former operates with only passive depth cues, such as parallax \cite{hirschmuller2005accurate,baek2016birefractive,meuleman2020single} and  defocus \cite{subbarao1994depth,subbarao1994depth}, to infer depth. These methods exhibit diminished accuracy for textureless scenes with few visual cues and complex geometries with ambiguous cues \cite{smolyanskiy2018importance}.
Active methods overcome this challenge by sending out a known illumination pattern into the scene and using the returned signal to help reconstruct depth. While structured light approaches rely on this illumination to improve local image contrast \cite{scharstein2003high, ahuja1993active}, ToF imaging uses the travel time of light itself to measure distance~\cite{heide2015doppler,shrestha2016computational}. This sensing approach allows for compact illumination-sensor setups and does not hinge on ambiguous visual cues.

\vspace{0.5em}\noindent\textbf{ToF Imaging.}\hspace{0.1em}
ToF imaging can be further categorized into direct and indirect methods. Direct ToF devices such as LiDAR send out pulses of light, scanning over a scene and directly measuring their round-trip time via avalanche photodiodes~\cite{cova1996avalanche,pandey2011ford} or single-photon detectors~\cite{mccarthy2009long,heide2018sub}. While accurate and long-ranged, these systems can produce only a few spatial measurements at a time, resulting in sparse depth maps~\cite{ma2019self}. Furthermore, their specialized detectors are orders of magnitude more expensive than conventional CMOS sensors. AMCW ToF imaging, a representative indirect ToF method, instead floods the whole scene with periodically modulated light and infers depth from phase differences between captures~\cite{hagebeuker20073d, lange2001solid}. These captures can be acquired with a standard CMOS sensor, making AMCW ToF cameras an affordable solution for dense depth measurement. Ultimately, all these devices integrate light over an aperture and are thus susceptible to FPs~\cite{reynolds2011capturing,sabov2008identification}.

\vspace{0.5em}\noindent\textbf{Depth Reconstruction Methods.}\hspace{0.1em}
Depth cameras are all subject to erroneous measurements, which has led to a wide array of work in robust depth reconstruction algorithms. Some approaches attempt to learn a direct mapping between noisy and clean 3D points~\cite{mattei2017point,nurunnabi2015outlier}, though they are limited in their scope and scalability as they contend with graph operations on unstructured point cloud data~\cite{qi2017pointnet, qi2017pointnet++}. Correlation between color and depth has also been used to smooth noisy depth estimates and enforce view consistency~\cite{kopf2007joint,lindell2018single}, though these approaches often blur object edges, producing more FPs. Confidence-based methods for ToF~\cite{reynolds2011capturing, frank2009denoising} on the other hand can detect FPs as unreliable measurements, but lack the context needed to determine if they belong to an object, background, or intermediate depth. Mask-ToF resolves this problem with a two-stage, generalizable approach that joins reconstruction with the optical pipeline; where a spatially varying amplitude mask encodes the information needed to \textit{correct} these flying pixels.

\vspace{0.5em}\noindent\textbf{Masks for Computational Imaging.}\hspace{0.1em}
Masks enable an imaging system to directly modify the point spread function (PSF) of input light, densely encoding information about the scene that can be computationally recovered post-capture. Amplitude masks can only attenuate light, yet have a wide range of applications including light-field~\cite{marwah2013compressive}, lensless~\cite{asif2016flatcam}, x-ray~\cite{proctor1979design}, high-speed~\cite{llull2013coded}, and spectral imaging~\cite{arce2013compressive}. Phase masks can allow for finer manipulation of PSFs~\cite{colburn2018metasurface}, and may be of interest in future masked ToF projects, but are prohibitively expensive to manufacture at a micro-scale resolution. In this work we learn an occlusion mask with spatially varying microlens apertures, encoding scene geometric information in AMCW ToF measurements to help correct FPs during reconstruction.

\vspace{0.5em}\noindent\textbf{End-to-End Design of Optics and Computation.}\hspace{0.1em}
Conventional imaging systems are designed in a sequential manner: first develop the optical and sensor stack in isolation, driven by compartmentalized metrics, then delineate an image processing pipeline~\cite{tseng2019hyperparameter}. Recently, a new paradigm of jointly optimizing optics and reconstruction has emerged, where all stages are jointly optimized in the design phase. These hold promise for applications in extended depth-of-field \cite{sitzmann2018end}, microscopy \cite{nehme2020deepstorm3d}, monocular depth \cite{chang2019deep}, HDR \cite{sun2020learning}, hyperspectral \cite{baek2020end}, and transient~\cite{sun2020end} imaging. Inspired by these works, Mask-ToF uses a differentiable ToF simulator to jointly learn an optimal mask pattern and train a depth refinement network to produce high SNR, low FP depth maps.

%% file: 3image_formation.tex
\section{Image Formation}
%\subsection{Conventional ToF}
Before introducing our proposed method, we review the fundamentals of AMCW ToF imaging; for details see~\cite{lange20003d}.

\vspace{0.5em}\noindent\textbf{Pinhole Model.}\hspace{0.1em} 
Correlation ToF cameras flood-illuminate the scene with an amplitude-modulated light signal
\begin{equation}\label{eq:light_source}
  p(t) := \alpha \cos(\omega t) + \beta.
\end{equation}
Here $\omega$ is a modulation frequency, $\alpha$ is amplitude, and $\beta$ is signal bias. Under a \textit{pinhole} camera model this modulated light is perfectly reflected by an object and captured by the ToF camera after travel time $\tau$. The measured signal
\begin{equation}\label{eq:detected}
  \tilde{p}(t - \tau) = \tilde{\alpha} \cos(\omega t - \phi) + \tilde{\beta}, \quad \phi = \omega \tau
\end{equation}
is effectively $p(t)$ with attenuated amplitude $\tilde{\alpha}$, shifted bias $\tilde{\beta}$, and an introduced $\tau$-dependent phase shift $\phi$. \vfill\eject\noindent The camera then correlates $\tilde{p}(t - \tau)$ with an identically modulated reference signal $r(t)=\cos(\omega t + \psi)$ to produce
\begin{equation}\label{eq:correlation}
C ( \psi ) = \int_{0}^{T} {\tilde p( {t - \tau } )r(t) dt} \;\approx\; \frac{\tilde \alpha}{2}\cos(\phi + \psi), \\
\end{equation}
for integration time $T\gg\tau$. By sampling correlation values $C(\psi)$ at four different phase offsets $\psi\!=\!\left[0,\, \pi/2,\, 3\pi/2,\, \pi\right]$, we can extract the measured signal's true phase $\phi$ from
\begin{equation}\label{eq:depth}
  \phi = \arctan\left(\frac{C(\pi)-C(\pi/2)}{C(0)-C(3\pi/2)}\right) + 2\pi n.\quad  n \in \mathbb{N}.
\end{equation}
This arctangent, however, introduces a $2 \pi n$ phase ambiguity for depths $z\geq\lambda=c/2\omega$, halved for the round-trip distance and with $c$ being the speed of light. To estimate this factor $n$ we can use a phase unwrapping algorithm~\cite{lawin2016efficient}, which typically solves instances of Equation~\eqref{eq:depth} for multiple modulation frequencies $\omega$ and disambiguates $\phi$ via Euclidean division~\cite{xia2007phase}. This estimate is ultimately converted to depth as $z=\phi c/4\pi\omega$.

\begin{figure}[t]
    \centering
    \includegraphics[width=\linewidth]{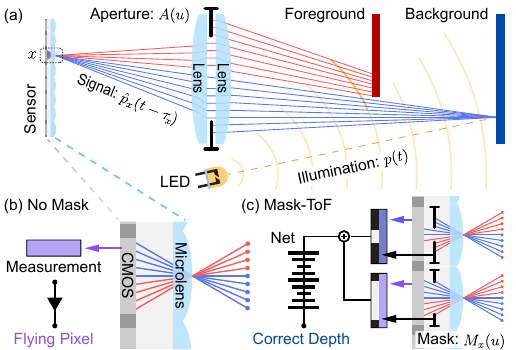}
    \caption{\label{fig:flying_pixels}%
     (a) ToF measurement at an object edge. (b) Without a mask, signals mix in unknown quantities to produce a flying pixel. (c) Mask-ToF can use surrounding pixel values and mask structure to disambiguate this measurement. 
    }
    \vspace{-3mm}
\end{figure}

\vspace{0.5em}\noindent\textbf{Lens Model.}\hspace{0.1em}
In a practical camera system, to increase light throughput, we use a lens to focus light incident on an aperture plane $\mathcal{U}$ to a sensor pixel $x$; for simplicity we assume a 2D model. We thus rewrite the image formation model as
\begin{equation}\label{eq:detected_ideal_lens}
  \hat{p}_x (t - {\tau_x}) = \int_{u \,\in \, \mathcal{U}} A(u) \tilde{p}_x(t - {\tau_x}-\tau_u) du,
\end{equation}
where $\hat p_x$ is the measurement at pixel $x$, $A(\cdot)$ is a binary aperture function, $u\in\mathcal{U}$ is the aperture coordinate, and $\tau_u$ is an additional time-of-flight term incurred by the residual path length. $\tilde p_x, \tau_x$ refer to Equation~\eqref{eq:detected} evaluated for a ray connecting through point $u$ to $x$.
The length of this ray is
\begin{align}\label{eq:aperture_distance}
d_{u,x} = \sqrt{(r-u)^2 + f^2} + \sqrt{z^2 + (\frac{xz}{f} + u)^2},
\end{align}
where $f$ is the focal length of the lens, $z$ is the depth of the scene point, and $r$ is the lens radius.
For a typical AMCW ToF camera, with operating range $z \gg r$ and modulation frequency $\omega = \mathcal{O}(10^8\mathrm{Hz})$, the phase contribution from the residual distance term $\delta_{u,x} = d_{u,x} - d_{0,x}$ is negligible. We thus discard the corresponding time-of-flight term $\tau_u$, and approximate the image formation model as
\begin{equation}\label{eq:detected_ideal_lens}
  \hat p_x (t - {\tau_x}) \approx \int_{u \,\in \, \mathcal{U}} A(u) p_x(t - {\tau_x}) du.
\end{equation}

\vspace{0.5em}\noindent\textbf{Flying Pixels.}\hspace{0.1em}
While for an unobstructed point $z$ this image formation is adequate, an edge case arises for points at a depth discontinuity. Suppose there is a single pixel on the sensor whose chief ray ($u=0$) comes from an $x$ near an object edge, see Figure~\ref{fig:flying_pixels}. This would mean that for part of the aperture coordinates, $\mathcal{U}^F$, we would receive unfocused light rays from the foreground object, with travel time $\tau'$, while the other rays passing through $\mathcal{U}^B=\mathcal{U} \setminus \mathcal{U}^F$ would have the intended travel time $\tau$. The received signal would similarly consist of a mix of both foreground $\hat{p}^F_x(t - \tau'_x)$ and background $\hat{p}^B_x(t - \tau_x)$ measurements
% \begin{equation}
\begin{align}\label{eq:detected_real}
&\hat p(t - \tau) := \hat p^F(t - \tau') + \hat p^B(t - \tau)  \nonumber\\
\Rightarrow \: &\hat p(t - \tau) = \tilde{\alpha} \cos(\omega t - \phi) + \tilde{\alpha}' \cos(\omega t - \phi') + \tilde{\beta} + \beta', \nonumber\\
&\phi = \omega\tau,\; \phi' = \omega\tau' \nonumber\\
\Rightarrow \: &\hat{\phi} = \arctan\left(\frac{\tilde\alpha \sin(\phi) + \tilde\alpha' \sin(\phi')}{\tilde\alpha \cos(\phi) + \tilde\alpha' \cos(\phi')}  \right),
\end{align}
where $\hat{\phi}$ is the measured phase shift of this mixed signal. Solving Equation~\eqref{eq:depth} returns an incorrect depth $\hat{z}$ somewhere between the foreground and background depths.

\vspace{0.5em}\noindent\textbf{Aperture-Masked ToF Image Formation.}\hspace{0.1em}
It might seem that a simple solution to the above flying pixel problem is just to reduce the aperture size. In the extreme case where $A(u)=0,\forall u>0$, we retain only the chief ray and so have no mixed measurements. Unfortunately, this also leads to poor light efficiency, which lowers the system's SNR as it becomes more susceptible to photon shot. We provide a detailed discussion of this fundamental SNR/FP tradeoff in the Supplemental Document. To better maintain light throughput, we can selectively block light paths by applying a spatially-varying microlens amplitude mask $M_x(u)$ to the image plane. The model from Equation (\ref{eq:detected_ideal_lens}) thus becomes
\begin{equation}\label{eq:mask}
\hat p_x(t - \tau_x) =\int_{u \,\in \, \mathcal{U}} M_x(u)A(u)p_x(t - {\tau_x}) du.
\end{equation}
One could imagine an \textit{ominscient} mask $M_x(u)=0$ for $u \in \mathcal{U}^F$, else $M_x(u)=1$. This would remove unfocused foreground light and preserve all other light paths, perfectly correcting FPs with high SNR. Unfortunately, such a mask could only work for a single scene, and we would need to know that scene beforehand to design it. Instead, with the derivations above, we can form a differentiable framework for AMCW ToF image formation and use gradient descent to learn a single generalizable mask pattern. We describe this approach in the following section.

%% file: 4method.tex
\begin{figure*}[t]
    \centering
    \includegraphics[width=\textwidth]{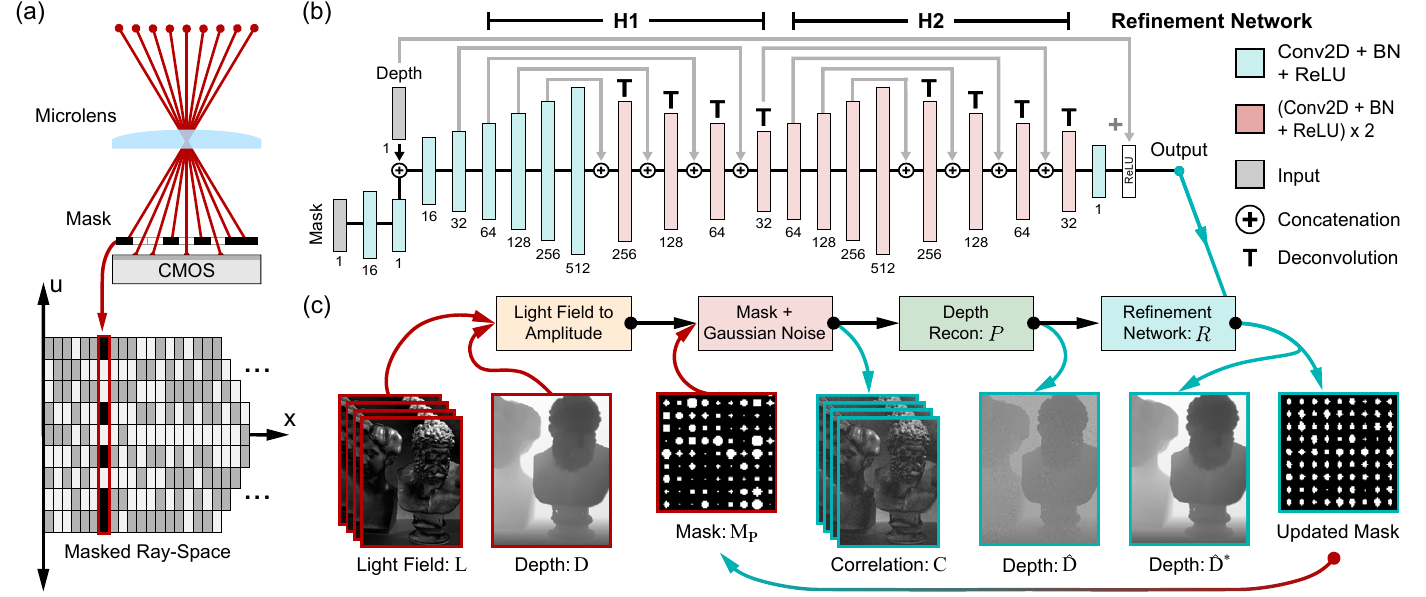}
    \caption{\label{fig:main_fig}%
    (a) Our microlens mask enables spatially-varying aperture coding in the angular-spatial domain.
    (b) The refinement network $R$, with skip connections in gray, which aggregates mask and coded depth information to produce a reduced FP output.
    (c) Stages of the differentiable simulation pipeline, from input light-field data to output refined depth.}
    \vspace{-4mm}
\end{figure*}

\section{Learning to Mask Flying Pixels}

\vspace{0.5em}\noindent\textbf{Mask Intuition.} Before we outline how to learn a mask, it's important to intuit why a static mask could help correct FPs in the first place. With a global aperture, shown in Figures \ref{fig:snr_fp_mini} and \ref{fig:flying_pixels} (b), all pixels are equally susceptible to FPs; if one sensor pixel returns an FP, likely so will its neighbors. The addition of spatially variable susceptibility via a microlens mask, shown in Figures \ref{fig:snr_fp_mini} and \ref{fig:flying_pixels} (c), means this is no longer the case. A sensor pixel with a wide effective aperture can be trusted with regards to noise statistics but is likely to return FPs when near an object boundary. Contrastingly, a neighboring pixel with a narrow aperture will likely produce noisier measurements, but be less affected by depth discontinuities. By aggregating information in pixel neighborhoods, we can effectively use wide aperture pixels to denoise local measurements, and narrow aperture pixels to de-flying-pixel them. This means a Mask-ToF approach critically needs not only a mask, but also a method to decode the information encoded by the mask.

 \vspace{0.5em}\noindent\textbf{From Light to Time-of-Flight.}\hspace{0.1em}
Given ground truth depth, we can simulate ToF measurements via Equation~\eqref{eq:detected} without distinguishing between light rays. However, to apply the mask $M_x(u)$ as in Equation~\eqref{eq:mask}, we also need access to the aperture plane $\mathcal{U}$. We thus discretize the image formation model and use light fields~\cite{levoy1996light} as a natural parametrization 
\begin{equation}\label{eq:mask_discrete}
\hat p_{u,x}(t - \tau_x) =\sum_{u \, \in \, L} M_x(u)A(u)p_x(t - {\tau_x}).
\end{equation}
Here $\hat p_{u,x}$ is our ToF signal, a sum over sub-aperture views $u$ in the light field $L$, discretized now in both $x$ and $u$. As the number of sub-aperture views $|L|\rightarrow \infty$ we converge on the form of Equation~\eqref{eq:mask}, though in practice $|L|$ governs mask resolution and is limited by manufacturing constraints.

\vspace{0.5em}\noindent\textbf{Tensor Image Model.}\hspace{0.1em} Rather than operate on $\hat p_{u,v,x,y}$, in 3D space we swap to a tensor view of simulation, as visualized in Figure~\ref{fig:main_fig} (c). We start with a depth map $\mathrm{D}\in\mathbb{R}^{H {\times} W}$, which we convert to phase array $\Phi$, and with the light field tensor $\mathrm{L}$ simulate $4 {\times} |\mathrm{L}|$ correlation images $C_{0,0}-C_{3,|L|}$. One for each of four phase shifts $\psi$ and sub-aperture arrays $\mathrm{L}_u \in \mathrm{L}$. These are individually masked by $\mathrm{M}_u$, and the views are averaged to produce 4 final correlation images $\mathrm{C}_\psi$, subject to simulated noise $\eta_\psi$. This process is summarized in Equation \eqref{eq:recon+noise}. \pagebreak
\begin{align}\label{eq:recon+noise}
    &\mathrm{C}_{\psi, u} = \mathrm{L}_u \odot (0.5 + \cos(\Phi + \psi))\frac{gT}{\pi}, \quad \Phi=\frac{4\pi\omega}{c}\mathrm{D} \nonumber \\
    &\mathrm{C}_{\psi} = \eta_\psi + \frac{1}{|\mathrm{L}|}\sum_{u} \mathrm{M_{u}} \odot \mathrm{C}_{\psi,u} \nonumber\\
    & \eta_\psi \sim \mathrm{uniform}(a,b) \cdot \mathcal{N}^{H {\times} W}(\mu, \sigma^2).
\end{align}
Here $g$ is sensor gain, $T$ is integration time, $H {\times} W$ is the sensor size, and $\odot$ denotes element-wise multiplication. The noise constants $a,b,\mu,\sigma$ are chosen empirically. At high photon counts, Poisson and Skellam \cite{callenberg2017snapshot} noise can be well approximated by scaled Gaussian noise, thus $\eta$ generalizes many expected sources of ToF noise~\cite{hansard2012time} while maintaining simulation differentiability.

\begin{figure*}[t]
    \centering
    \includegraphics[width=\linewidth]{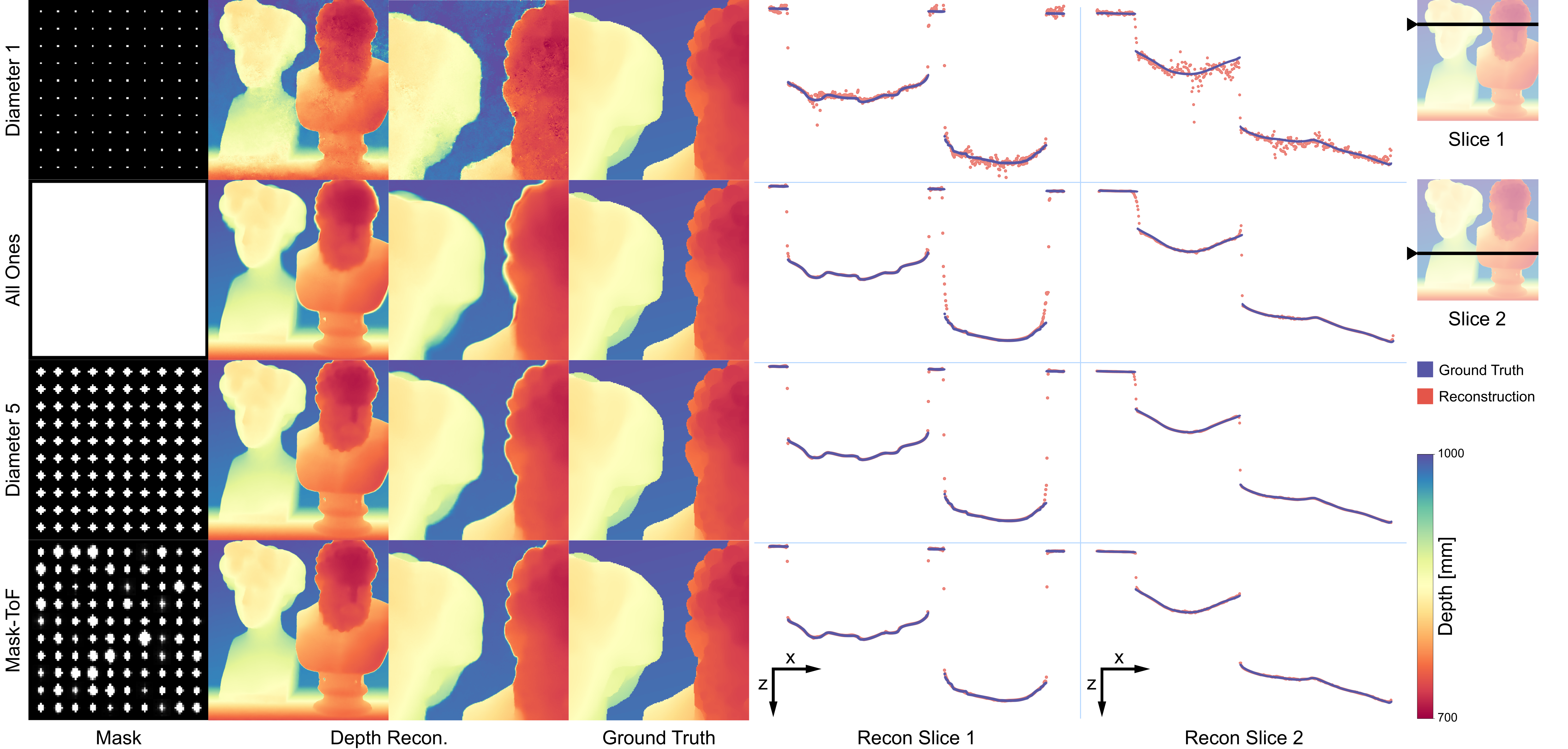}
	\caption{\label{fig:comparisons}%
	Comparison of optimized and naïve mask design results on simulated test data, with zoomed view of reconstruction and ground truth regions. Slices of the depth maps are plotted to help visualize flying pixels and noise susceptibility.
    }
    \vspace{-5mm}
\end{figure*}

\vspace{0.5em}\noindent\textbf{Depth Reconstruction.} Using Equation~\eqref{eq:depth} we generate an estimated depth map $\mathrm{\hat D}$ from the four masked correlation images $\mathrm{C}_\psi$. We implement this as a differentiable function $\mathrm{\hat D} = P(\mathrm{C})$ with automatic gradient evaluations. This grants us flexibility as we can swap $P(\mathrm{C})$ for other depth estimation methods, such as the discrete Fourier transform, if needed. To process the information embedded in these measurements by the microlens mask, we propose a refinement network $R$, illustrated in Figure \ref{fig:main_fig} (b). $R$ is a residual encoder-decoder model, inspired by the hourglass architecture from \cite{xu2020aanet}, which takes as input $\mathrm{\hat D}$ and $\mathrm{M}$ and outputs
\begin{equation}\label{eq:res_network}
\mathrm{\hat D^*} = R(P(\mathrm{C}), \mathrm{M})=\max(0,\mathrm{\hat D} + \mathrm{\hat D^R}),
\end{equation}
\noindent where $\mathrm{\hat D^*}$ is the refined depth map and $\mathrm{\hat D^R}$ is a learned residual depth which when added to $\mathrm{\hat D}$ serves to correct the now spatially multiplexed effects of noise and FPs. As Equation~\eqref{eq:depth} does the initial depth calculation, $R$ does not have to learn how to generate depth from phase, and can be made significantly more lightweight than a typical deep reconstruction network. This helps $R$ to quickly learn high-level depth and mask features,    and generalize well to arbitrary scenes where raw phase data might significantly differ from the training set. The sequential depth estimation and refinement approach also allows us to naturally exploit calibration procedures~\cite{kolb2010time} implemented by the sensor manufacturer. We can feed real depth data directly into $R$ without having to retrain and learn calibration offsets.

\vspace{0.5em}\noindent\textbf{Loss Functions.}\hspace{0.1em}
For training, we opt for a combined loss
\begin{align}\label{eq:loss}
    &\mathcal{L} = \frac{1}{HW} \sum_i \left( w_L \mathcal{L}_{S}(\mathrm{\hat D^*_i}, \mathrm{D_i}) + w_C \mathcal{L}_{C} (\mathrm{\hat D^*_i}, \mathrm{D}) \right) \nonumber , \\
    &\mathcal{L}_{S}(\mathrm{\hat D^*_i}, \mathrm{D_i}) = \left\{\begin{array}{ll}
    |\mathrm{\hat D^*_i} - \mathrm{D_i}| - \delta/2 & \text { if } |\mathrm{\hat D^*_i} - \mathrm{D_i}|\geq\delta \\
    (\mathrm{\hat D^*_i} - \mathrm{D_i})^2/2\delta & \text { else },
    \end{array}\right. \nonumber \\
    &\mathcal{L}_{C}(\mathrm{\hat D^*_i}, \mathrm{D}) = \min_j \| \mathrm{proj}(\mathrm{\hat D^*_i}) - \mathrm{proj}(\mathrm{D_j}) \|,
\end{align}
where $i,j\in\{0,...,HW-1\}$ are enumerative indices. Smooth L1 loss $\mathcal{L}_{S}$ helps enforce local smoothness in the reconstructed depth map, controlling the Gaussian noise $\eta$ while being less sensitive to depth outliers. To penalize these outliers, we add a Chamfer loss term $\mathcal{L}_{C}$. It considers the projected points $\mathrm{proj}(p)$, which we produce by concatenating sensor coordinates $x,y$ to the corresponding depth values $z$, and penalizes points based on their distance to the nearest ground truth point. FPs, which exist in the empty space between foreground and background depths with no close neighbors, are thus heavily penalized. We balance these losses with weights $w_L$ and $w_C$. \pagebreak

\vspace{0.5em}\noindent\textbf{Patch-based Training.}\hspace{0.1em}
We can propagate the gradient of this loss through the differentiable framework above all the way to the mask $\mathrm{M}$, however, learning an unconstrained $\mathrm{M}$ proves computationally burdensome. We thus restrict $\mathrm{M}$ to be an $m {\times} n$ tiling of a mask patch $\mathrm{M_P}$ and learn $\mathrm{M_P}$ instead, adopting a patch-based stochastic gradient descent approach. We sample a batch of random patches from the input light field and ground truth depth, and update $\mathrm{M_P}$ and $R$ based on the patch loss. This effectively exponentially increases the number of available samples for training, allowing us to rely on a relatively small light field dataset. As $R$ is fully convolutional, it is invariant to input shape, and so this patch training generalizes well to the full-sized images.

%% file: 5synthetic_assessment.tex
\section{Synthetic Assessment} \label{sec:eval}
\vspace{0.5em}\noindent\textbf{Implementation.}\hspace{0.1em}
Our network $R$ is trained on patches of size $80{\times}80{\times}9{\times}9$ sampled from $512{\times}512{\times}9{\times}9$ synthetic light field data. This data, sourced from \cite{honauer2016benchmark}, contains $9{\times}9$ sub-aperture views per image pixel, and a total of 16 light fields. The full-sized evaluation masks are constructed by an $8{\times}8$ tiling of the center $64{\times}64$ area of the learned mask $\mathrm{M_P}$, reducing edge artifacts from training. Noise parameters $a,b,\mu,\sigma$ from Equation~\eqref{eq:recon+noise} are set to 0.75, 1.25, 0, 3; empirically matched to real recorded ToF samples.
We set sensor gain to $g=20$ and integration time to $T=1$\,ms.

The ADAM optimizer was used for training~\cite{kingma2014adam}, with an initial learning rate of $0.004$ for the refinement network $R$ and $0.1$ for the mask $\mathrm{M_P}$. We halve both rates every 80 epochs. $\mathrm{M_P}$ is not updated for the first 70 epochs of training, as these epochs tend to be extremely noisy, the convolutional layers of $R$ having not yet learned high-level structures~\cite{zeiler2014visualizing}. Through empirical study we found weights $w_L=100$ and $w_C=0.08$ to effectively balance the differing scales of Chamfer and smooth L1 loss in Equation~\eqref{eq:loss}. We leave $\delta$ as the default $\delta=1$. The network contains 19 million learnable parameters, 1 million of which are the amplitude mask, and is trained for 3 hours on a single Nvidia Tesla V100. Inference time for a single 512${\times}$512 image is $\approx 8$\,ms. Code and trained models will be made available.

\vspace{0.5em}\noindent\textbf{Ablation Study.}\hspace{0.1em}
We quantify the effects of architecture design choices in a series of ablation experiments, summarized in Table~\ref{table:ablation}. Here, \textit{Proposed} is our final network $R$, and \textit{Chamfer Only}$/$\textit{L1 Only} are tests where we train $R$ using only the respective loss function. In the \textit{Half} modification we remove the first hourglass, \textbf{H1}, while in \textit{Big} we double intermediate channel counts in $R$. These help gauge if the network can be simplified or requires increased parametrization. The \textit{Global} modification adds another global channel, implemented by duplicating \textbf{H1} with increased stride lengths and concatenating the new signal at the input to \textbf{H2}, to test if the network can be improved by aggregating more non-local information. The \textit{No Mask} tests the effect of training $R$ on only depth data, without mask input. Lastly, \textit{ToFNet} is a reimplementation of the ToFNet architecture from \cite{su2018deep}. We train it until convergence with weighted L1 and TV loss as suggested in the original work, and fine-tune the learning rate to our data.
\begin{table}[t]
	\resizebox{\columnwidth}{!}{
	\begin{tabular}{ c c  c  c  c }
			%\hline
			\toprule
			Ablation & RMSE $\downarrow$ & MAE $\downarrow$ & Thresh 3mm $\downarrow$ & Thresh 15mm $\downarrow$\\
			\midrule
			\midrule
			\textbf{Mask-ToF}       &\textbf{5.166}$/$\textbf{7.115}&\textbf{1.281}$/$\textbf{1.278}&5.052$/$\textbf{4.397}&\textbf{1.178}$/$\textbf{1.120}\\
			\textbf{Chamfer Only}   &6.459$/$7.913&1.992$/$1.930&10.91$/$9.878&1.457$/$1.330\\
			\textbf{L1 Only}        &5.216$/$7.127&1.284$/$1.293&\textbf{5.024}$/$4.426&1.214$/$1.194\\
			\textbf{Half}           &5.489$/$7.432&1.356$/$1.367&5.247$/$4.647&1.391$/$1.373\\
			\textbf{Big}            &5.432$/$7.169&1.514$/$1.482&6.351$/$5.439&1.369$/$1.284\\
			\textbf{Global}         &5.427$/$7.310&1.407$/$1.393&5.488$/$4.716&1.363$/$1.307\\
			\textbf{No Mask}        &5.482$/$7.353&1.410$/$1.398&5.284$/$4.664&1.369$/$1.303\\
			\textbf{ToFNet}	        &11.42$/$12.19&5.120$/$5.038&42.36$/$42.82&5.316$/$4.964\\
			\bottomrule
			
	\end{tabular}
	}
	\vspace{-3mm}
	\caption{\label{table:ablation}%
	Quantitative ablation results (train$/$test) for changes to network $R$ or training procedure. \textit{Thresh \textbf{X}mm} is a threshold metric denoting the percentage of points further than \textbf{X} millimeters from ground truth depth.
	}
	\vspace{-3mm}
\end{table}

\begin{table}[t]
	\resizebox{\columnwidth}{!}{
	\begin{tabular}{ c c  c  c  c }
		\toprule
		Mask & RMSE $\downarrow$ & MAE $\downarrow$ & Thresh 3mm $\downarrow$ & Thresh 15mm $\downarrow$\\
			\midrule
			\midrule
		Diam. 1    &9.412$/$8.293&5.203$/$4.576&46.31$/$46.549&6.647$/$4.345\\
		%\hline
		All Ones   &9.227$/$12.58&2.470$/$2.814&9.712$/$10.45&3.118$/$3.558\\
		%\hline
		Diam. 5    &6.512$/$8.732&1.718$/$1.753&7.377$/$6.552&1.585$/$1.582\\
		%\hline
		Mask-ToF  &\textbf{5.166}$/$\textbf{7.115}&\textbf{1.281}$/$\textbf{1.278}&\textbf{5.052}$/$\textbf{4.397}&\textbf{1.178}$/$\textbf{1.120}\\
			\bottomrule
			
	\end{tabular}
	}
	\vspace{-3mm}
	\caption{\label{table:quantitative_comparison}%
	Quantitative comparison (train$/$test) of mask-aided ToF recovery. 4 images (\textit{greek}, \textit{pillow}, \textit{pens}, \textit{tower}) of 16 withheld for testing. 
	}
	\vspace{-5mm}
\end{table}

We validate the proposed architecture of the network $R$ with the results in Table~\ref{table:ablation}. Specifically, the proposed method wins in all categories compared to the \textit{Big} and \textit{Half}, suggesting it is adequately parametrized. The lack of improvement from \textit{Global} also suggests that the network $R$ is sufficiently utilizing non-local information. \textit{Chamfer Only} and \textit{No Mask} both lead to lackluster performance, emphasizing the value of the L1 regularization term and mask comprehension, respectively. Although we see close results for \textit{L1 Only}, the addition of Chamfer loss does lead to a reduction in outliers, expressed in RMSE and threshold metrics. \textit{ToFNet} shows overall worse performance than our \textit{Baseline} refinement architecture, with the network learning to reconstruct a smooth depth map, however not learning to remove flying pixels. This is possibly due to its significantly wider scope; lacking a skip layer to the output, it must learn to reconstruct depth from raw phase measurements.

\vspace{0.5em}\noindent\textbf{Analysis of Mask Patterns.}\hspace{0.1em}
Mask-ToF contains a feedback loop: a change in the mask structure of $\mathrm{M_P}$ necessitates an update to the refinement network $R$, which itself alters the propagated loss gradient and changes the structure of $\mathrm{M_P}$. Thus, to avoid local minima, we test a broad set of both human-selected and randomly generated initial masks including: various diameters of circular aperture, Gaussian and Bernoulli noise, randomly oriented \textit{barcode} structures, and several multiplexed designs. A full discussion of mask patterns is available in the Supplemental Document.

We compare the final optimized mask against the best hand-crafted (naïve) initializations to validate our proposed end-to-end optimization method. For a fair comparison, we fine-tune the refinement network $R$ for each of these hand-crafted mask designs and highlight the drop in performance from a lack of joint mask optimization. Results are displayed in Figure \ref{fig:main_fig} and quantified in Table \ref{table:quantitative_comparison}. We see that the \textit{Diameter 1} mask achieves low error for the 15mm threshold metric and RMSE, which we find to be a good proxy for FP count.
Even with the refinement network, however, its low light throughput leads to a large amount of noise, resulting in poor MAE and RMSE values. On the other end of the spectrum, the \textit{All-Ones} (open-aperture) mask produces smooth low-noise reconstructions, but with copious FPs. The optimized mask design wins in all categories, with the low FP count and high SNR, and provides near-identical light throughput as the \textit{Diameter 5} mask (13 times the throughput of the \textit{Diameter 1} pinhole mask).

%% file: 6experimental_assessment.tex
\section{Experimental Assessment}

\begin{figure}[t]
    \centering
    \includegraphics[width=\columnwidth]{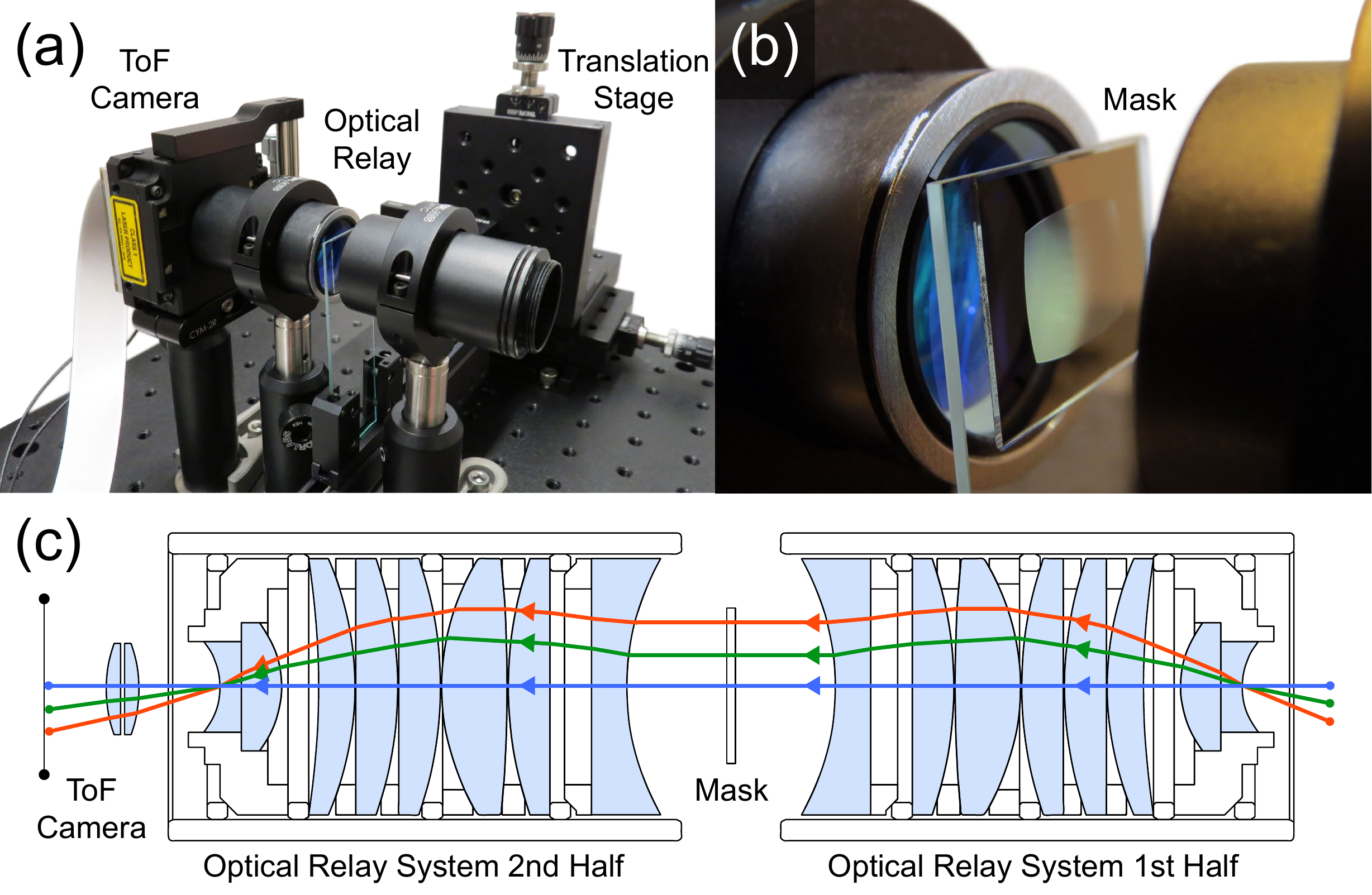}
    \caption{\label{fig:rig}%
    (a) The assembled imaging system.
    (b) The mask mounted on a precision microscope slide attached to a translation stage.
    (c) A schematic of the relay lens system.
    }
    \vspace{-5mm}
\end{figure}

\vspace{0.5em}\noindent\textbf{Mask Fabrication.}\hspace{0.1em}
We fabricate our custom mask patterns via photolithography.
0.5mm fused silica wafers are used as the substrate, receiving a 200nm of chromium film to occlude light. A layer of 0.6 $\mu$m thick photoresist AZ1505 is then spin-coated on top. We place the wafer under a master mask on a contact aligner (EVG 6200$\infty$) for UV exposure, and develop in AZ726 to form the mask pattern on the photoresist. With an etchant we then remove the chromium from under open areas in the photoresist. See the Supplemental Document for further information on fabrication.

\begin{figure*}[t]
    \centering
    \includegraphics[width=\linewidth]{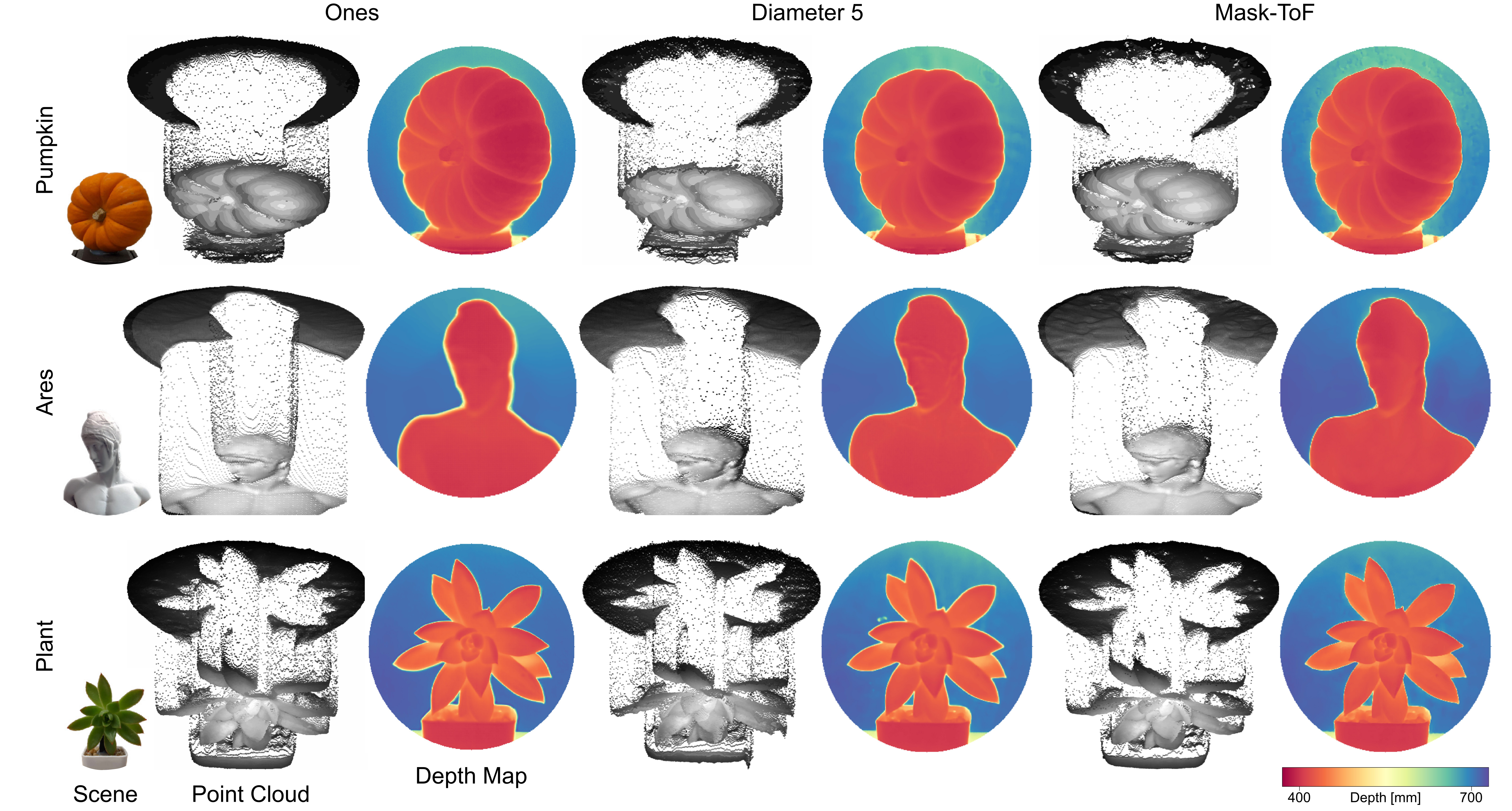}
    \caption{\label{fig:qual_real}%
    Perspective point cloud visualizations and depth maps of reconstruction results for Mask-ToF and naïve mask designs; object references on the left. Point cloud texture is generated from separate long-exposure amplitude captures.
    }
    \vspace{-5mm}
\end{figure*}

\begin{figure}[t]
    \centering
    \includegraphics[width=\columnwidth]{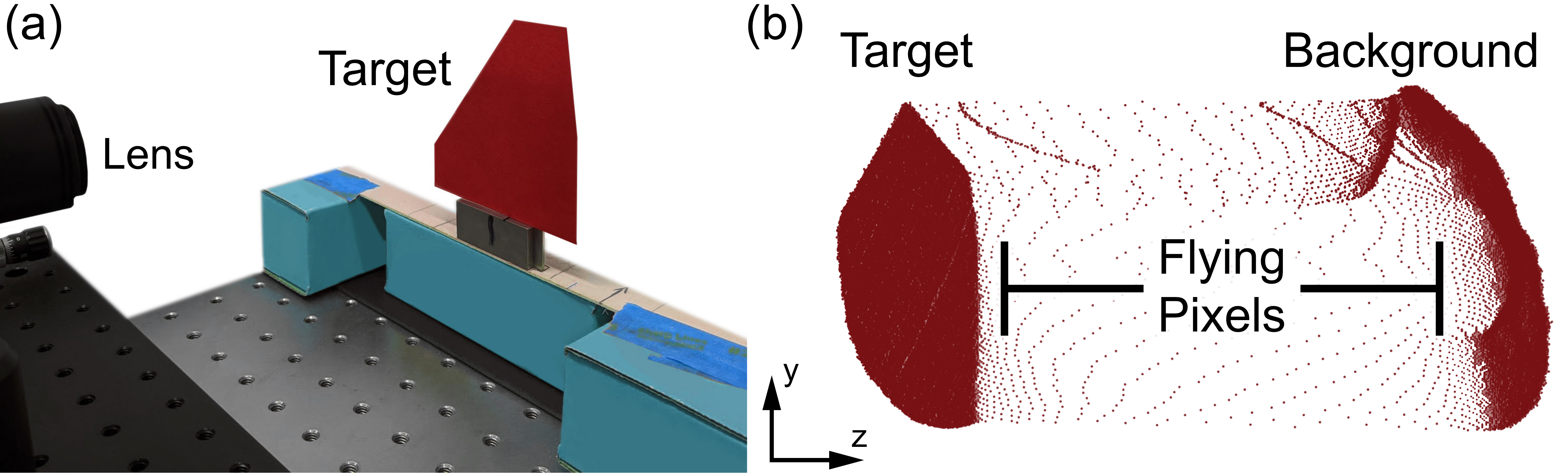}
    \begin{tabular}{c c c c}
         & Ones & Diameter 5 & Mask-ToF \\
        \midrule
        \midrule
        Flying Pixel Ratio: & 1 & 0.69 & 0.48 \\

    \end{tabular}
    \caption{\label{fig:quant_real}%
    To quantify flying pixel ratios, we (a) capture 5 flat targets at a known depth and (b) count the number of points between the target and background planes.
    }
    \vspace{-1em}
\end{figure}
\vspace{0.5em}\noindent\textbf{Prototype.}\hspace{0.1em}
We capture measurements with an AMCW ToF camera (Helios Flex, Lucid Vision) operating on an NVIDIA Jetson TX2. We use a custom-designed 1:1 Keplerian telescope as an optical relay system to virtually place the mask on the sensor (see Figure~\ref{fig:rig}). This eliminates the need to remove the sensor cover glass and allows for rapid prototyping, but in a commercial product can be supplanted by a directly integrated mask to maintain device form factor. The mask sits on the intermediate image plane of the telescope attached to a precision microscope slide, which is optically conjugate with the sensor plane. We adjust the position of the mask with an XYZ translation stage (Thorlabs PT3A). For more details, see the Supplemental Document.\pagebreak

\vspace{0.5em}\noindent\textbf{Results.}\hspace{0.1em}
Depth maps were captured via the previously described setup and fed directly into the synthetically trained refinement network $R$, with no network fine-tuning. By counting points as in Figure~\ref{fig:quant_real}, we see that Mask-ToF cuts flying pixel counts in half when compared to an open aperture. Compared to the near-identical light throughput \textit{Diameter 5} mask we reduce FPs by an additional 30.5$\%$. These results are qualitatively confirmed in Figure~\ref{fig:qual_real} for objects of varying geometry and reflectance, with additional results in the Supplemental Document and 3D rendering in the accompanying video. Our optimized mask reconstruction visibly and significantly reduces FPs as compared to \textit{Diameter 5}, while maintaining object shape consistency with the open aperture measurements. Of note is how sharply Mask-ToF reconstructs the tips of the \textit{Plant} example's petals, as compared to the noisy reconstruction produced by the \textit{Diameter 5} mask. Additionally, Mask-ToF is even able to reduce intra-object FPs such as those inside the \textit{Plant}'s pot.

%% file: 7conclusion.tex
\section{Conclusion}
Mask-ToF is an end-to-end approach to tackle the long-standing problem of flying pixel artifacts in time-of-flight imaging. It learns a per-pixel microlens amplitude mask, that, when combined with a jointly trained refinement network, reduces FPs while preserving light throughput. We validate the method both in simulation and experimentally, manufacturing the learned mask and optically placing it on a camera sensor with a custom-designed optical relay system. The proposed mask and reconstruction method outperform existing hand-engineered masks (and no mask) for real-world scenes. In a mass-market implementation of our method, we envision the amplitude mask to be integrated as part of the sensor assembly, maintaining the camera form-factor while improving FP statistics. Future research directions include learned phase mask patterns and dynamic masks, implemented via a spatial light modulator or similar, which adapt their structure to the observed scene.

\vspace{0.5em}\noindent\textbf{Acknowledgements.}\hspace{0.1em} This work was supported in part by KAUST baseline funding, Felix Heide's NSF CAREER Award (2047359) and Sony Faculty Innovation Award, and Ilya Chugunov's NSF Graduate Research Fellowship.